\documentclass[a4paper, conference]{IEEEtran}
\IEEEoverridecommandlockouts
\usepackage{cite}
\usepackage{amsmath,amssymb,amsfonts,amsthm} 
\usepackage{amsmath, bm}
\usepackage{algorithmic}
\usepackage{algorithm}
\usepackage{graphicx}
\usepackage{textcomp}
\usepackage{xcolor}
\usepackage{color}
\usepackage{array}
\usepackage{ulem}

\newtheoremstyle{note}%
  {0pt}%
  {0pt}%
  {}%
  {1em}%
  {\itshape}%
  {:}%
  {0.5em}%
  {}%
\theoremstyle{note}
\newtheorem{theorem}{Theorem}
\newtheorem{lemma}{Lemma}
\newtheorem{remark}{Remark}

\def\BibTeX{{\rm B\kern-.05em{\sc i\kern-.025em b}\kern-.08em
		T\kern-.1667em\lower.7ex\hbox{E}\kern-.125emX}}

\begin{document}
\title{Minimizing End-to-End Latency for Joint Source-Channel Coding Systems
}

\author{\IEEEauthorblockN{Kaiyi Chi$^{*}$, Qianqian Yang$^{\dagger}$, Yuanchao Shu$^{*}$, Zhaohui Yang$^{\dagger}$, Zhiguo Shi$^{\dagger}$}
    \IEEEauthorblockA{
        $^{*}$College of Control Science and Engineering, Zhejiang University, Hangzhou 310027, China\\
        $^{\dagger}$College of Information Science and Electronic Engineering, Zhejiang University, Hangzhou 310027, China\\
        E-mail: \{kaiyichi, qianqianyang20, ycshu, yang\_zhaohui, shizg\}@zju.edu.cn}
    
    \thanks{This work is partly supported by NSFC under grant No. 62293481, No. 62201505, partly by the SUTD-ZJU IDEA Grant (SUTD-ZJU (VP) 202102).}
}

\maketitle
\begin{abstract}
   While existing studies have highlighted the advantages of deep learning (DL)-based joint source-channel coding (JSCC) schemes in enhancing transmission efficiency, they often overlook the crucial aspect of resource management during the deployment phase. In this paper, we propose an approach to minimize the transmission latency in an uplink JSCC-based system. We first analyze the correlation between end-to-end latency and task performance, based on which the end-to-end delay model for each device is established. Then, we formulate a non-convex optimization problem aiming at minimizing the maximum end-to-end latency across all devices, which is proved to be NP-hard. We then transform the original problem into a more tractable one, from which we derive the closed form solution on the optimal compression ratio, truncation threshold selection policy, and resource allocation strategy. We further introduce a heuristic algorithm with low complexity, leveraging insights from the structure of the optimal solution. Simulation results demonstrate that both the proposed optimal algorithm and the heuristic algorithm significantly reduce end-to-end latency. Notably, the proposed heuristic algorithm achieves nearly the same performance to the optimal solution but with considerably lower computational complexity.
\end{abstract}



\section{Introduction}
Existing communication systems are developed based on Shannon's separation theorem, in which source coding and channel coding are separate steps. Source coding focuses on eliminating source redundancy, while channel coding introduces redundant information to enhance resilience against channel noises. While this separation is theoretically optimal with an infinitely large block length in memory-less channels \cite{shannon}, practical implementations often involve finite block lengths. Recently, the emergence of deep learning (DL) has prompted researchers to explore joint source-channel coding (JSCC) schemes using DL techniques. DL-based JSCC models designed for text \cite{deepsc}, image \cite{jscc}, speech \cite{speech} transmission, etc., have demonstrated significant advantages over their separated counterparts.

Traditional communication systems primarily focus on maximizing the transmission quality like bit error rate (BER) in resource allocation. However, this approach is not suitable for JSCC systems, where the emphasis is on the application performance rather than the quality of bit transmission. Despite this, the majority of research efforts focus on developing Deep Learning (DL)-based JSCC models, often overlooking the resource allocation challenge during system deployment. Some strides have been taken to address resource allocation strategies in DL-based JSCC systems in existing literature \cite{performance, resource, NOMA}.
In \cite{performance}, the authors tackled the resource allocation problem in downlink text transmission. They maximized the defined metric of semantic similarity (MSS) by jointly optimizing the transmission of semantic information and selecting resource blocks. In \cite{resource}, the authors introduced the semantic transmission rate (S-R) and semantic spectral efficiency (S-SE), proposing to maximize the overall S-SE in an uplink scenario. Lastly, in \cite{NOMA}, the authors delved into the resource allocation challenges in the downlink non-orthogonal multiple access (NOMA) system.

The aforementioned works focus on optimizing the weighted task performance of all devices. However, in some latency sensitive scenarios, we prefer the end-to-end delay of the device to be as small as possible. In addition, with the increasing size of the neural networks and the limited computation resources, the computational latency should also be taken into consideration.
Motivated by these considerations, this paper aims to minimize the maximum end-to-end latency of the uplink transmission from all devices in the system while ensuring task performance. 

We begin by theoretically analyzing the delay model to understand the relationship between task performance and end-to-end latency, which yields an end-to-end delay model for each device. We then formulate an optimization problem aimed at minimizing the maximum latency across all devices while ensuring task performance. This involves jointly considering the selection of compression ratio, channel truncation threshold, and the allocation of communication and computation resources. Recognizing the NP-hard nature of the problem, we employ problem transformation to make it more tractable. We obtain a closed-form solution for the optimal compression ratio, channel truncation threshold selection strategy, and resource allocation policy. Furthermore, we propose a heuristic algorithm with low complexity to tackle the problem in practical considerations. Simulation results validate the effectiveness of the proposed methods in terms of minimizing the end-to-end uplink latency.

The rest of the paper is organized as follows. Section \uppercase\expandafter{\romannumeral2} introduces the system model and formulates the problem. Section \uppercase\expandafter{\romannumeral3} presents the proposed solution, followed by simulation results
in Section \uppercase\expandafter{\romannumeral4}. Section \uppercase\expandafter{\romannumeral5} concludes the paper.

\section{System Model and Problem Formulation}
In this section, we first introduce the DL-based JSCC system and establish the end-to-end latency model which takes both the communication and computation latency into account. Based on this, we then formulate an optimization problem to minimize the maximum latency among all devices while guaranteeing the task performance of each device.
	
\subsection{System Model}
\begin{figure}[htbp]
    \vspace{-8ex}
    \centerline{\includegraphics[width=3in]{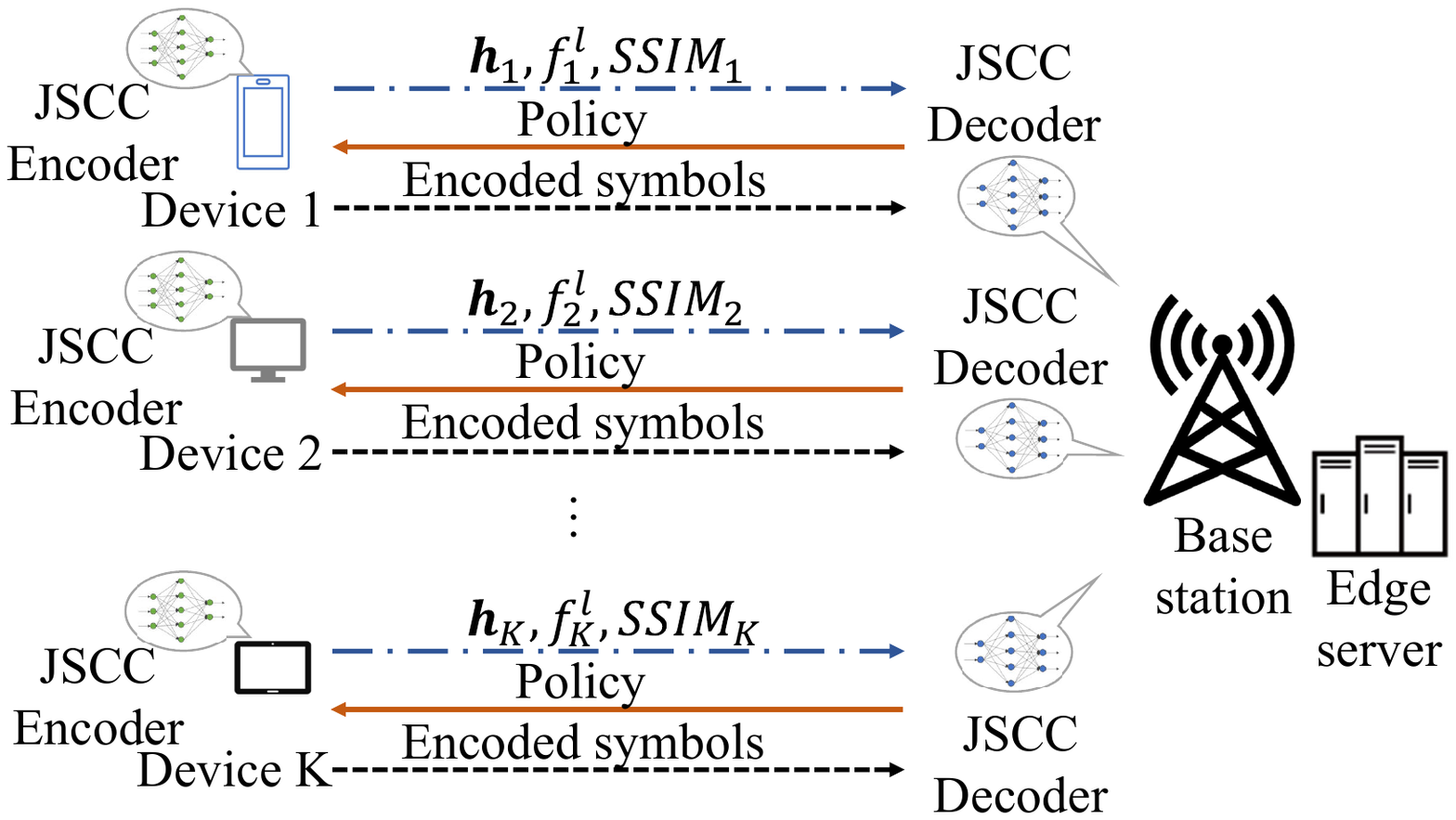}}
    \vspace{-5ex}
    \caption{The considered JSCC system model.}
    \vspace{-3ex}
    \label{fig-system}
\end{figure}

Consider an uplink cellular network with a base station (BS) and a set $\mathcal{K}=\{1,2,...,K\}$ of $K$ devices, as shown in Fig.~\ref{fig-system}. Each device aims to accomplish a specific task. We assume image transmission in this paper for the simplicity of analysis. However, we note that our proposed method can be extended into other transmission tasks, such as text transmission or speech transmission. We adopt the convolutional
neural networks (CNN) based joint source-channel coding (DeepJSCC) network proposed in \cite{jscc}, where the encoder is deployed at the local device and the decoder is deployed at the edge server connected with the BS. We note that it can be extended to other types of networks such as DNN with similar analysis. 
The considered system operates as follows. Each device sends its information about channel state information (CSI), performance constraint, as well as the available local computation resource to the BS. Then, the BS will determine the communication and computation resource allocation policy of each device. After that, each device will compress the image with a specific JSCC network and transmit the extracted symbols to the BS via physical channel. Then, the BS decodes the content of received symbols using corresponding decoder network in parallel. 

\subsection{Encoding at Local Device}
Each device uses an encoder locally to compress its source image of size $D_{0}=3\times H \times W$, where $H$ and $W$ are the height and width, respectively. We define the compression ratio $o_{k}$ as the proportion of the number of the transmitted symbols versus to the total number of symbols in the input image of device $k$. We have $N$ predefined compression ratios (CRs), i.e., $o_{k} \in \{c_{1}, c_{2},..., c_{N}\}, \forall k \in \mathcal{K}$, and each CR corresponds to a specific encoder and decoder. 
We denote $C_{k}^{l}$ as the computational cost per image of device $k$ during local processing.
We assume the images of all devices have the same resolution, i.e., $H$ and $W$. According to \cite{cnn}, the computational cost of a CNN is proportional to the size of input resolution, i.e., $HW$. 
Thus, the computational complexity of the encoder at local device $k$ is $L_{k}C_{k}^{l}=L_{k}C^{s}HW$, where $L_{k}$ is the number of images to be processed at device $k$. $C^{s}$ is the required number of CPU cycles per pixel using the encoder, which is determined by the architecture of the encoder network, i.e., CR value. We note that the $C^{s}$ is almost the same for different encoders with different CRs in our test since they only differ in the number of feature maps at the output layer. Therefore, $C^{s}$ is obtained through average under different CRs. 
Given that the local CPU-cycle frequency at the local device $k$ is $f_{k}^{l}$, the computational latency of the encoding process at device $k$ is 
\vspace{-1ex}
\begin{equation}
    t_{k}^{l}=\frac{L_{k}C_{k}^{l}}{f_{k}^{l}}, \forall k \in \mathcal{K}.
\end{equation}

\subsection{Transmission Model}
We assume that time-division multiple access (TDMA) is applied for the channel access. Hence, each time frame is slotted and we denote the time slot allocated to device $k$ for transmission per unit time as $\tau_{k}$.
We assume orthogonal frequency division multiplexing (OFDM) modulation where the whole bandwidth $B$ is divided into $M$ orthogonal sub-channels. At $t$-th time-slot, the received symbol from device $k$ at the $m$-th sub-carrier is given by \cite{over-the-air}
\begin{equation}
    y_{k}^{m}(t)=r_{k}^{-\frac{\alpha}{2}} h_{k}^{m}(t) p_{k}^{m}(t) x_{k}^{m}(t)+z(t),
\end{equation}
where $r_{k}^{-\frac{\alpha}{2}}$ denotes the path-loss of the link between the BS and device $k$, $r_{k}$ denotes the distance between them and $\alpha$ is the path-loss exponent. $h_{k}^{m}(t)$ denotes the small scale fading of the channel following Rayleigh fading $\mathcal{C} \mathcal{N}(0,1)$, which is identically and independently distributed (i.i.d.) over $k, m, t$. $p_{k}^{m}(t)$ is the power allocated to the device $k$ on the $m$-th sub-carrier over $t$-th time slot. $z(t)$ is the Gaussian channel noise with power of $\sigma^2$.
For ease of notation, we will omit the index $t$ in the following.

Following \cite{over-the-air}, we let the power allocated to each sub-carrier $p_{k}^{m}$ adapt to the channel coefficient $h_{k}^{m}$ to achieve the signal-to-noise ratio (SNR) alignment at the BS. We assume each device is subject to a long-term transmission power constraint
$\mathbb{E}\left[\sum_{m=1}^{M}\left|p_{k}^{m}\right|^{2}\right] \leq P_{k}, \forall k \in \mathcal{K},$
where $P_{k}$ is the maximum transmission power of the device $k$.
Since channel coefficients are identically and independently distributed over different sub-channels, the above power constraint can be reformulated as
\begin{equation}
    \mathbb{E}\left[\left|p_{k}^{m}\right|^{2}\right] \leq \frac{P_{k}}{M}, \forall k \in \mathcal{K}.
    \label{power constraint}
\end{equation}

We assume the knowledge of perfect CSI at each device. The device can thus perform power control on each sub-carrier to allow the received signals at the BS have the same amplitude across different subcarriers. Besides, to cope with deep fades, we adopt a more practical truncated channel inversion. To be more specific, a sub-channel will be cutoff for a device at this time slot if its channel coefficient is less than a threshold $g_{k}$, i.e.,  
\begin{equation}
    \begin{aligned}
        p_{k}^{m}=\left\{\begin{array}{ll}
            \frac{\sqrt{\rho_{k}}}{r_{k}^{-\frac{\alpha}{2}} h_{k}^{m}}, & \left|h_{k}^{m}\right|^{2} \geq g_{k}, \\
            0, & \left|h_{k}^{m}\right|^{2} < g_{k},
        \end{array}\right.
        \label{power allocation}
    \end{aligned}
\end{equation}
where $\rho_{k}$ is a scaling factor in order to meet the power constraint, and its also the power of the received symbols transmitted by the device $k$.

Since the channel coefficients follow Rayleigh distribution $\mathcal{C} \mathcal{N}(0,1)$, the channel gain of the $k$-th link on the $m$-th sub-carrier $\left|h_{k}^{m}\right|^{2}$ follows the exponential distribution with unit mean. 
With truncated power allocation, we have
$\mathbb{E}\left[\left|p_{k}^{m}\right|^{2}\right] =
\frac{\rho_{k}}{r_{k}^{-\alpha}} \int_{g_{k}}^{\infty} \frac{1}{g} \exp (-g) d g \leq \frac{P_{k}}{M}.$
Thus, the maximum received power of the symbols transmitted by the $k$-th device is bounded by
\begin{equation}
    \rho_{k} \leq \frac{P_{k}}{M r_{k}^{\alpha} \operatorname{Ei}\left(g_{k}\right)}, \forall k \in \mathcal{K},
    \label{received power}
\end{equation}
where $\mathrm{Ei}(g_k)=\int_{g_k}^{\infty} \frac{1}{g} \exp (-g) d g$.

Besides the receive SNR, channel truncation ratio is also affected by the power-cutoff threshold $g_{k}$. 
We denote the percentage of the channels that are not truncated as the activate ratio $\zeta_{k}$. 
When the number of transmitted symbols of device $k$ is large enough, the activate ratio is equal to the probability that the channel gain is above the power-cutoff threshold, i.e., $\zeta_{k} 
    =\operatorname{Pr}\left(\left|h_{k}\right|^{2}>g_{k}\right)
    =e^{-g_{k}}, \forall k \in \mathcal{K}.$
Hence, the number of expected activate channels per time slot is $M e^{-g_{k}}$. Denote the symbol duration of an OFDM symbol by $T_{s}$. Thus, the transmission delay of device $k$ is given by
\begin{equation}
    t_{k}^{t}=\frac{D_{0} o_{k}}{M e^{-g_{k}}} \frac{T_{s}}{\tau_{k}}, \forall k \in \mathcal{K}.
    \label{delay}
\end{equation}

\subsection{Decoding at the BS}
The BS decodes the messages transmitted from the devices in parallel. We denote that the total computation resource of the edge server by $F^{c}$, 
and $f_{k}^{c}$ as the computation resource allocated to decode the message from device $k$, which satisfies $\sum_{k} f_{k}^{c} \leq F^{c}$. Similarly, we denote $C_{k}^{d}$ as the computational cost to decode an image at the decoder of device $k$, and $C^{s'}$ is the computational cost per pixel. Thus, the computational complexity for decoding message from device $k$ at the decoder is $L_{k}C_{k}^{d}=L_{k}C^{s'}HW$. Thus, the computational latency of decoding message from device $k$ at the edge is given as
\begin{equation}
    t_k^c=\frac{L_kC_k^d}{f_k^c}, \forall k \in \mathcal{K}.
\end{equation}
Hence, the end-to-end latency of encoding, transmitting, and then decoding message from device $k$ can be given as $t_{k}=t_{k}^{l}+t_{k}^{t}+t_{k}^{c}, \forall k \in \mathcal{K}$.

\subsection{Performance Metrics}
\begin{figure}[htbp]
    \vspace{-5ex}
    \centerline{\includegraphics[width=2.7in]{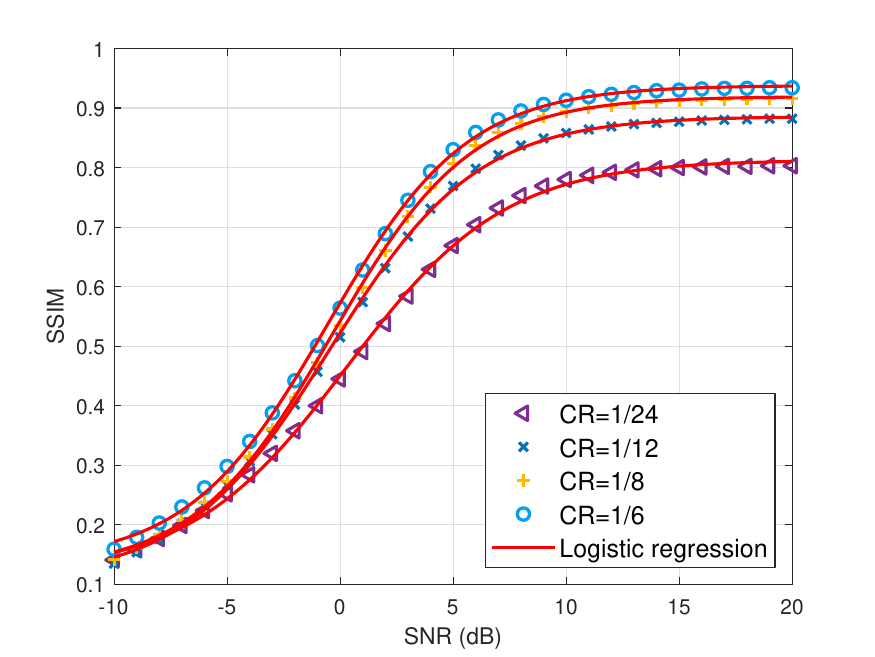}}
    \vspace{-2ex}
    \caption{Average SSIM of the reconstructed images vs. SNR under different compression ratios on ImageNet dataset.}
    \vspace{-3ex}
    \label{fig-ssim}
\end{figure}

In the considered system, the BS needs to take the performance on the uplink transmission of each device into consideration. We adopt the structure-similarity-index-measure (SSIM) as the performance metrics to evaluate the considered image transmission task as it can capture the perceived visual quality of the images well. According to the \cite{capacity optimization}, the SSIM of the reconstructed image is determined by both the SNR of the received signal and the compression ratio. The SSIM by the adopted JSCC scheme increases as the compression ratio and SNR as shown in Fig.~\ref{fig-ssim}, where the simulation settings are shown in Section \uppercase\expandafter{\romannumeral4}. We then use a generalized logistic function to fit the curve, which is given by

\begin{equation}
    SSIM(o, \gamma)=A_{o,1}+\frac{A_{o,2}-A_{o,1}}{1+e^{-(C_{o,1}\gamma+C_{o,2})}}, 
    \label{ssimk}
\end{equation}
where $\gamma$ is the SNR of the device, and $A_{o,1}$, $A_{o,2}$, $C_{o,1}$, $C_{o,2}$ are constant values when the compression ratio $o$ is given. 


\subsection{Problem Formulation}
In this paper, we aim at minimizing the maximum end-to-end latency of the uplink transmission among all devices, which we refer to as the system delay. We formulate this problem as follows:
\begin{subequations}
    \begin{align}
        \mathcal{P}1:\  \min _{\{o_{k}, g_{k}, \tau_{k}, f_{k}^{c}\}} & \max 
_{k \in \mathcal{K}} \  t_{k} \label{p1a} \\
        \text { s.t. } 
        & \operatorname{SSIM}_{k} \geq \eta_{k}, \forall k \in \mathcal{K}, \label{p1b} \\
        & \sum_{k=1}^{K} \tau_{k} \leq 1, \label{p1c} \\
        & \sum_{k=1}^{K} f_{k}^{c} \leq F^{c}, \label{p1d} \\
            & g_{k} \geq 0, \forall k \in \mathcal{K}, \label{p1e} \\
        & o_{k} \in\left\{c_{1}, c_{2}, \ldots, c_{N}\right\}, \forall k \in \mathcal{K}, \label{p1f} \\
        & \tau_{k} \geq 0, f_{k}^{c} \geq 0, \forall k \in \mathcal{K}, \label{p1g}
    \end{align}
\end{subequations}
where $\eta_{k}$ denotes the performance requirement of each device, and \eqref{p1b} ensures the SSIM requirement is met for each device. 
\eqref{p1c} limits the overall communication resource of all devices, 
\eqref{p1d} ensures the overall computation resource allocated to decode the message for each device can not exceed the threshold,
\eqref{p1e} is the truncation threshold constraint to ensure that it is a non-negative value, \eqref{p1f} limits the range of the compression ratio within a given set. 
It can be observed that $\mathcal{P}1$ is a mixed integer non-linear problem (MINLP), which is hard to solve. 
In the next section, we will develop an effective algorithm to solve this problem.

\section{Proposed Solution}
In this section, we first transform the $\mathcal{P}1$ into an equivalent problem, then we propose an efficient method to address the transformed problem. 

\subsection{Problem Transformation}
We reformulate the original problem into a more tractable one by introducing an auxiliary variable $T$ as follows \cite{delay}.

\begin{subequations}
    \begin{align}
        \mathcal{P}2: \min_{\{T,\tau_{k},g_{k},o_{k},f_{k}^{c}\}} & T \\
        s.t. & \frac{L_{k}C_{k}^{l}}{f_{k}^{l}}+\frac{L_{k}D_{0}o_{k}}{Me^{-g_{k}}}\frac{T_{s}}{\tau_{k}}+\frac{L_{k}C_{k}^{d}}{f_{k}^{c}}\leq T, \forall k \in \mathcal{K}, \label{p2b} \\
        & \mathrm{\eqref{p1b}} - \mathrm{\eqref{p1g}}.
    \end{align}
\end{subequations}

Let $\{T^*,\tau_k^*,g_k^*,o_k^*,f_k^{c^*}\}$ be the optimal solution to the problem. We then obtain the following lemma.

\begin{lemma}
Solution to $\mathcal{P}2$ with $T<T^*$ is infeasible while the solution with $T>T^*$ is feasible. 
\label{lemma-1}
\end{lemma}

\begin{IEEEproof}
Please refer to Appendix \ref{proof lemma 1}.
\end{IEEEproof}

Based on the Lemma 1, we provide the algorithm to obtain the solution of $\mathcal{P}2$ by bisection search, as presented in Algorithm~\ref{alg:algorithm1}. Specifically, we now derive the upper bound and lower bound of $T$, i.e., $T_{max}$ and $T_{min}$, to initialize the searching space. Intuitively, any feasible solution to the problem $\mathcal{P}2$ with a $T^{\prime}$ can be viewed as the upper bound, since the optimal delay is no more than this solution, i.e., $T^* \leq T^{\prime}$. For instance, we can equally allocated communication and computation resource to all devices, i.e., $\tau_{1}=...=\tau_{K}=1/K, f_1^{c}=...=f_K^{c}=F^{c}/K$, and we can randomly choose available $g_k,o_k$ to obtain the upper bound.
Regarding the lower bound of $T$, we can assume that there is just one device $k$ to be served and all the resources are allocated to the device, and $g_k,o_k$ are optimized respectively.

It is not straight forward to test the feasibility of $\mathcal{P}2$ when $T$ is fixed. Instead, we transform $\mathcal{P}2$ into an equivalent problem $\mathcal{P}3$ as follows. It is evident that the solution to $\mathcal{P}2$ is feasible if the objective function value of the $\mathcal{P}3$ is less than 1 to satisfy the communication resource allocation constraint in \eqref{p1c}.


\begin{subequations}
    \begin{align}
        \mathcal{P}3: \min_{\{\tau_{k},g_{k},o_{k},f_{k}^{c}\}} & 
 \sum_{k \in \mathcal{K}}\tau_{k} \\
        s.t.
        & \frac{L_{k}C_{k}^{l}}{f_{k}^{l}}+\frac{L_{k}D_{0}o_{k}}{Me^{-g_{k}}}\frac{T_{s}}{\tau_{k}}+\frac{L_{k}C_{k}^{d}}{f_{k}^{c}}\leq T,\forall k \in \mathcal{K}, \label{p3b} \\
        & \operatorname{SSIM}_{k} \geq \eta_{k}, \forall k \in \mathcal{K}, \label{p3c} \\
        & \sum_{k=1}^{K}f_{k}^{c}\leq F^{c}, \label{p3d} \\
        & g_{k}\geq0,\forall k \in \mathcal{K}, \label{p3e} \\
        & o_{k}\in\{c_{1},c_{2,}\ldots,c_{N}\},\forall k \in \mathcal{K}, \label{p3f} \\
        & \tau_{k}\geq0,f_{k}^{c}\geq0,\forall k \in \mathcal{K}. \label{p3g} 
    \end{align}
\end{subequations}

\vspace{-1ex}
\begin{algorithm}
    \caption{The Optimal Algorithm to $\mathcal{P}1$}
    \label{alg:algorithm1}
    \begin{algorithmic}[1] 
        \STATE {Initialize $T_{min}$, $T_{max}$, and the tolerance $\varepsilon$}
        \REPEAT
            \STATE {Set $T=(T_{max}+T_{min})/2$.}
            \STATE {Check the feasibility of the solution to $\mathcal{P}2$.}
            \STATE {If $\{T,\tau_{k}^{\prime},g_{k}^{\prime},o_{k}^{\prime},f_{k}^{c^{\prime}}\}$ is a feasible solution to the problem, set $T_{max}=T$, else, set $T_{min}=T$.}
        \UNTIL{$(T_{max}-T_{min})/T_{max} \leq \varepsilon$.}
    \end{algorithmic} 
\end{algorithm}
\vspace{-1ex}


\subsection{Optimal Solution}
In this subsection, we will discuss how to obtain the optimal solution to the $\mathcal{P}3$. We can first obtain the following theorem.

\begin{theorem}
$\mathcal{P}3$ is an NP-hard problem. 
\label{theorem-1}
\end{theorem}
\begin{IEEEproof}
    Please refer to Appendix \ref{proof theorem 1}. 
\end{IEEEproof}

To make it more tractable, we first focus on a simple scenario where the compression ratio of each device is fixed. Accordingly, the problem can be formulated as 

\begin{subequations}
    \begin{align}
        \mathcal{P}4: \min_{\{\tau_{k},g_{k},f_{k}^{c}\}}
        & \sum_{k \in \mathcal{K}}\tau_{k} \\
        s.t. 
        & g_k\geq d_k,\forall k \in \mathcal{K}, \label{p4c}\\
        & \mathrm{\eqref{p3b}}, \mathrm{\eqref{p3d}} - \mathrm{\eqref{p3g}}, 
    \end{align}
\end{subequations}
where $d_{k}$ can be obtained by solving $\int_{d_{k}}^{+\infty}\frac{1}{g}e^{-g}dg = c_{k}$ using bisection method, and $c_{k}$ is given as $c_k=\frac{P_k}{Mr_k^\alpha\sigma^2}10^{\frac{\ln\left(\frac{A_{o_k,2}-\eta_k}{\eta_k-A_{o_k,1}}\right)+C_{o_k,2}}{10C_{o_k,1}}}$ by substituting \eqref{received power} into \eqref{ssimk}.
We have the following theorem.

\begin{theorem}
$\mathcal{P}4$ is a convex optimization problem. 
\label{theorem-2}
\end{theorem}
\begin{IEEEproof}
    Please refer to Appendix \ref{proof theorem 2}.
\end{IEEEproof}

Based on Theorem \ref{theorem-2}, we can utilize the Lagrangian method to solve $\mathcal{P}4$. The partial Lagrangian function to $\mathcal{P}4$ can be given by

\begin{equation}
    \begin{aligned}
        \mathcal{L}=
        & \sum_{k \in \mathcal{K}}\tau_k+\sum_{k \in \mathcal{K}}\lambda_k(\frac{L_kC_k^l}{f_k^l}+\frac{L_kD_{0}o_k}{Me^{-g_k}}\frac{T_s}{\tau_k}+\frac{L_kC_k^d}{f_k^c}-T) \\
        & +\mu(\sum_{k \in \mathcal{K}}f_k^c-F^c),
    \end{aligned}
\end{equation}
where $\lambda_k$ and $\mu$ are the Lagrange multipliers associated with the constraints \eqref{p3b} and \eqref{p3d}, respectively. Let $f_{k}^{c^{*}}$, $\tau_{k}^{*}$ and $g_{k}^{*}$ denote the optimal solution to the $\mathcal{P}4$. Then, by utilizing the Karush-Kuhn-Tucker (KKT) conditions, we can derive the following theorem.

\begin{theorem}
The optimal solution to $\mathcal{P}4$ is given by
\begin{equation}
    f_{k}^{c^{*}}=\frac{L_{k}}{T-\frac{L_{k}C_{k}^{l}}{f_{k}^{l}}}(\sqrt{\frac{D_{0}o_{k}e^{g_{k}^{*}}T_{s}C_{k}^{d}}{M\mu^{*}}}+C_{k}^{d}), \forall k \in \mathcal{K}, \label{optimal_fkc}
\end{equation}

\begin{equation}
    \tau_{k}^{*}=\frac{L_{k}}{T-\frac{L_{k}C_{k}^{l}}{f_{k}^{l}}}(\frac{D_{0}o_{k}e^{g_{k}^{*}}T_{s}}{M}+\sqrt{\frac{\mu^{*}D_{0}o_{k}e^{g_{k}^{*}}T_{s}C_{k}^{d}}{M}}), \forall k \in \mathcal{K}, \label{optimal_tauk}
\end{equation}

\begin{equation}
    g_{k}^{*}=d_{k}, \forall k \in \mathcal{K}, \label{optimal_gk}
\end{equation}
where $\mu^{*}$ is the optimal Lagrange multiplier, as
\begin{equation}
    \mu^*=\left(\frac{\sum_k\frac{L_k}{T-{L_kC_k^l}/{f_k^l}}\sqrt{\frac{D_{0}o_ke^{g_k^*}T_sC_k^d}{M}}}{F^c-\sum_k\frac{L_kC_k^d}{T-{L_kC_k^l}/{f_k^l}}}\right)^2, \label{optimal_mu}
\end{equation}

\label{theorem-3}
\end{theorem}

\begin{IEEEproof}
    Please refer to Appendix \ref{proof theorem 3}.
\end{IEEEproof}

\begin{remark}
    We can find that the optimal allocated computing resource for device $k$ at BS
    , $f_{k}^{c^{*}}$, 
    increases with the local computational latency $\frac{L_{k}C_{k}^{l}}{f_{k}^{l}}$ according to \eqref{optimal_fkc}. Intuitively, if a device takes too much time on the local computation, the edge will allocate more computational resource to the device so as to reduce the end-to-end latency for the device such that to minimize the maximum latency among all devices. 
\end{remark}

So far, we have derived the optimal solution to $\mathcal{P}4$. Now we can obtain the algorithm to test the feasibility of problem $\mathcal{P}3$. One straight forward way is to exhaustively search the overall candidate compression ratio set $\{o_{1},...,o_{K}\}$ for devices and solve the corresponding $\mathcal{P}4$, then test whether the optimal objective value is smaller than 1. Thus we can obtain optimal solution to $\mathcal{P}1$ by this exhaustively search method. However, there are $N^{K}$ possible values for $\{o_{1},...,o_{K}\}$, which exhibits exponential complexity. 
In the sequel, we will propose a heuristic algorithm with low complexity.

\subsection{The Proposed Low-Complexity Algorithm}
In this subsection, we propose a heuristic algorithm to address the $\mathcal{P}1$ with low complexity. Recall the optimal $\tau_{k}^{*}$ in \eqref{optimal_tauk}, we note that the optimal $\tau_{k}^{*}$ is determined by $o_{k}e^{g_{k}}$. Moreover, since the objective of $\mathcal{P}3$ is the minimization of the sum of $\tau_{k}$, we can choose the compression ratio with the smallest $o_{k}e^{g_{k}}$ for each device individually. The detailed information of the heuristic algorithm is presented in Algorithm~\ref{alg:algorithm2}. The computational complexity of the bisection search of $T$ in the outer iteration is ${\mathcal O}(\log(^{1}/\varepsilon))$. Meanwhile, the computational complexity of choosing the compression ratio for each device is ${\mathcal O}(KN\log(^{1}/\varepsilon_{2}))$, where $\varepsilon_{2}$ is the error tolerance of the bisection method in computing $d_k$. Therefore, the overall computational complexity of the heuristic algorithm is ${\mathcal O}(\log(^{1}/\varepsilon)KN\log(^{1}/\varepsilon_{2}))$.

\begin{algorithm}
    \caption{The heuristic algorithm for $\mathcal{P}1$}
    \label{alg:algorithm2}
    \begin{algorithmic}[1] 
        \STATE {Initialize $T_{min}$, $T_{max}$, and the tolerance $\varepsilon$}
        \REPEAT
            \STATE {Set $T=(T_{max}+T_{min})/2$.}
            \FOR{Each device $k$}
                \FOR{$o_{k} \in\{c_{1},c_{2},\ldots,c_{N}\} $}
                    \STATE {Find the minimum $o_{k}e^{g_{k}}$ that satisfy the performance constraint.}
                \ENDFOR
            \ENDFOR
            \STATE {Solve the problem $\mathcal{P}4$ by Theorem \ref{theorem-2}, test whether the objective value is smaller than 1 or not. If true, set $T_{max}=T$, else, set $T_{min}=T$.}
        \UNTIL{$(T_{max}-T_{min})/T_{max} \leq \varepsilon$.}
    \end{algorithmic} 
\end{algorithm}
\vspace{-2ex}
\section{Simulation Results}
In this section, we present the simulations to demonstrate the performance of the proposed algorithm. The simulation parameters are set as follows unless otherwise stated. The BS has a coverage of 100 m, and we assume the path loss exponent $\alpha=3$, the number of sub-channels $M$ = 256, the bandwidth of each channel is 15 kHz, the noise variance is $\sigma^2=-80$ dBm. All devices have the same transmit power 0.1 W. The number of images for each device to transmit is uniformly generated from 1 to 10. The SSIM requirement for each device follows the uniform distribution within $\eta_{k} \in [0.8, 0.93]$. We adopt the DeepJSCC model in \cite{jscc} trained on the ImageNet \cite{imagenet} with a fixed SNR of 10 dB for training, and the SNR during testing varies from -10 to 20 dB. The size of the image is 128$\times$128. The optional compression ratio set is $\{1/6, 1/8, 1/12, 1/24\}$. We assume that the local CPU frequency of each device is uniformly distributed in [1, 2] GHz. The edge server is equipped with Intel(R) Core(TM) i7-11700F with 16 cores with 4.9 GHz per core. We use the \textit{cpulimit} \cite{cpulimit} to control the CPU usage of a process (denoted in percentage). Through simulation, we obtain that the computation cost per pixel for the encoders and decoders is about 2170 CPU cycles/pixel and 2510 CPU cycles/pixel in average, respectively.

The optimal solution derived in Section \uppercase\expandafter{\romannumeral3} B is referred to as \textit{OPT} and the heuristic algorithm proposed in Section \uppercase\expandafter{\romannumeral3} C is referred to as \textit{HEU}. For comparison, we consider three benchmarks. The first one equally allocates the communication and computation resources to each decoder, while the compression ratio and truncation threshold for each device are then optimized, named as \textit{EQU}. The second one equally allocates the computation and computation resources, and the compression ratios are fixed to the maximum for all devices while the truncation thresholds are optimized (ensure that the performance constraint can be satisfied), named as \textit{FIX\_O}. The third one also equally allocates the computation and computation resources, and the compression ratio is fixed to the maximum for each device, and the truncation threshold of each device is fixed to 0.5 (similarly, set the threshold to a high value to ensure that the performance constraint can be satisfied), named as \textit{FIX\_G}.

\begin{figure}[t]
    \centerline{\includegraphics[width=2.5in]{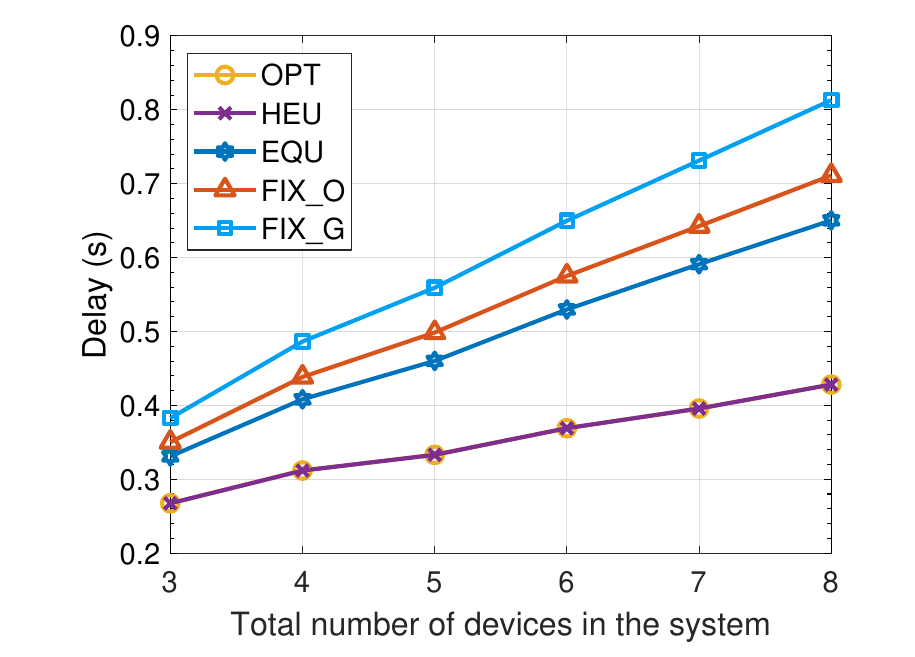}}
    \vspace{-2ex}
    \caption{Delay vs. number of devices.}
    \vspace{-5ex}
    \label{fig-devices}
\end{figure}

Fig.~\ref{fig-devices} shows the delay versus the number of devices. The computation resource at the edge server we use are two cores in total, which is denoted by 200\%. It can be seen that the delay increases with the number of devices for all schemes. We note that both the optimal method and the heuristic algorithm always outperform the benchmarks. We can observe that heuristic algorithm achieves almost the same performance as the optimal solution. The \textit{EQU} scheme shows performance degradation due to the fact that it can not utilize the communication and computation resource effectively. Besides, \textit{FIX\_O} performs even worse since it will transmit additional symbols when the performance constraint is low, which takes additional transmission time. Moreover, \textit{FIX\_G} degrades the performance even more because it deactivates too many channels when the performance constraint is low, which leads to a larger transmission delay.

\begin{figure}[htbp]
    \vspace{-4ex}
    \centerline{\includegraphics[width=2.5in]{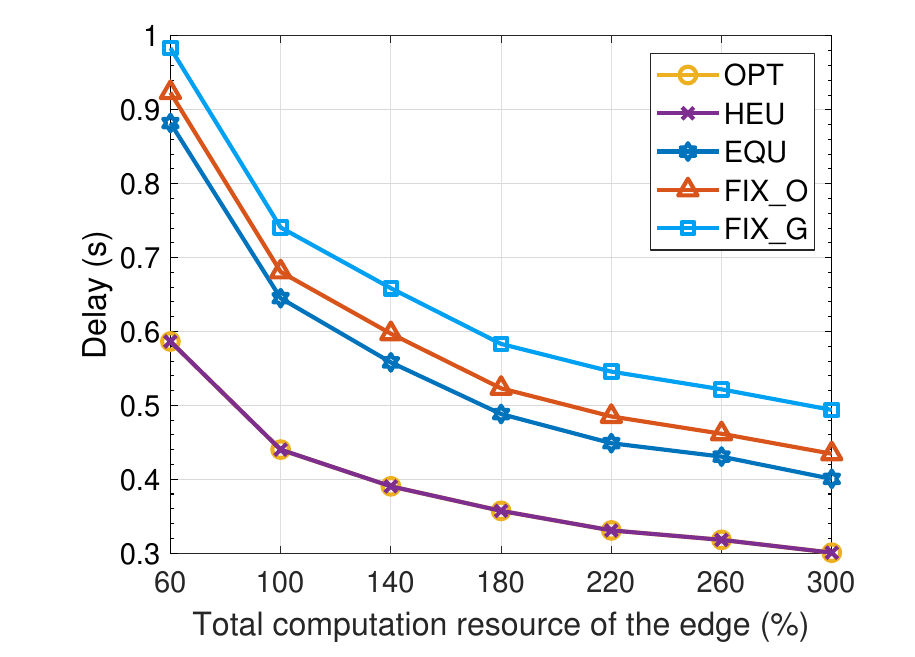}}
    \vspace{-2ex}
    \caption{Delay vs. edge computation resource. 
    (The percentage means the percentage of a CPU core, for example, 300\% means 3 CPU cores.)}
    \vspace{-3ex}
    \label{fig-edgecom}
\end{figure}

Fig.~\ref{fig-edgecom} depicts the system delay versus the edge computation resource, where the number of devices is set to 5. From the figure, we observe that the system delay decreases with the edge computation resource for all methods. Besides, both the optimal algorithm and heuristic algorithm outperform the benchmarks. 

\begin{figure}[t]
    \centerline{\includegraphics[width=2.5in]{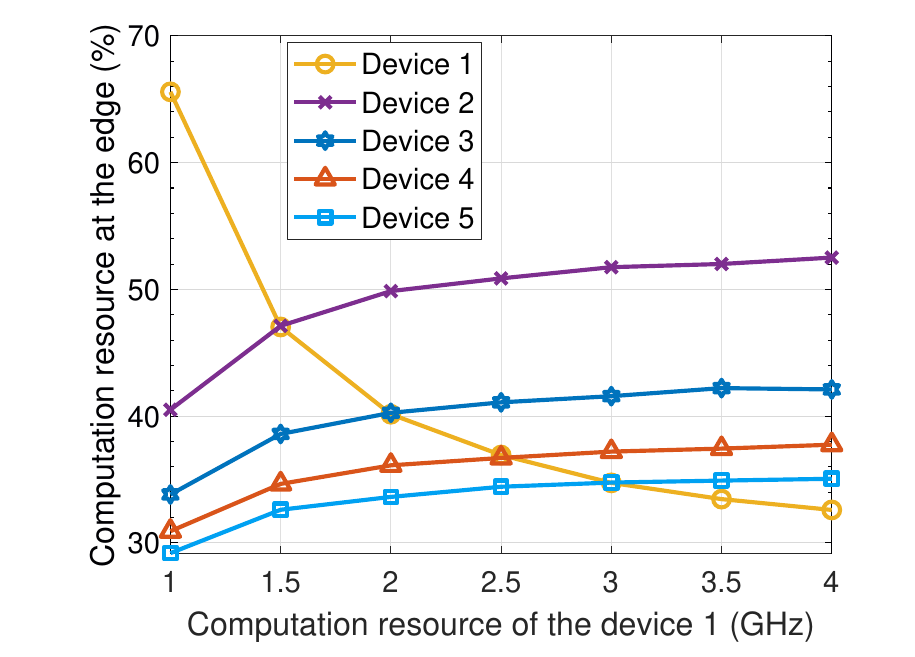}}
    \vspace{-2ex}
    \caption{Computation resource allocation vs. local computation resource of device 1. 
    (200\% in total, which means two CPU cores.)}
    \vspace{-3ex}
    \label{fig-dev1}
\end{figure}

Fig.~\ref{fig-dev1} shows the relationship between local computation resource and the edge computation resource. Here, all devices transmit 5 images to the BS. The computation resource at the device 1 varies from 1 to 4 GHz, while the computation resource at device 2, device 3, device 4, device 5 are fixed as 1.5 GHz, 2 GHz, 2.5 GHz, 3 GHz, respectively. As shown in the figure, with the increase of local computation resource at device 1, the percentage of edge computation resource allocated to decode message of device 1 decreases, while the edge computation resource allocated to other devices increases. This is intuitive due to the fact that the local computation time of device 1 will decrease with increasing local computation resource, thus less resource should be allocated to device 1 at the edge to ensure the fairness among devices, which is aligned with the optimal edge resource allocation policy in \eqref{optimal_fkc}.

\vspace{-1ex}
\section{Conclusion}
In this paper, we proposed a resource allocation scheme to minimize the end-to-end latency of the uplink DL-based JSCC systems. 
We analyzed the relationship between the end-to-end delay and the task performance of each device and then formulated the latency optimization problem, which is NP-hard. 
Through the problem transformation, we derived the closed form solution to the optimal compression ratio and channel truncation threshold selection policy and resource allocation strategy. Then we proposed an effective heuristic algorithm to solve the original problem with low computational complexity. Finally, simulation results demonstrated that both the proposed optimal algorithm and the heuristic algorithm can reduce end-to-end latency significantly. Remarkably, the proposed heuristic algorithm achieved nearly the same performance to the optimal solution but with much lower complexity. 

\begin{appendices}
\small

\vspace{-2ex}
\section{Proof of Lemma 1}
\label{proof lemma 1}
We give a proof by contradiction to the lemma. Assume that $\{T^{\prime},\tau_{k}^{\prime},g_{k}^{\prime},o_{k}^{\prime},f_{k}^{c^{\prime}}\}$ is a feasible solution with $T<T^*$. Thus, the solution has a lower objective function value $T^{\prime}$ than the optimal objective function value $T^*$, which contradicts to the assumption that $\{T^*,\tau_k^*,g_k^*,o_k^*,f_k^{c^*}\}$ is the optimal solution. Meanwhile, for the solution with $T^{\prime}>T^*$, we can construct a solution $\{T^{\prime},\tau_k^*,g_k^*,o_k^*,f_k^{c^*}\}$ which is always feasible. Thus, the proof completes.

\vspace{-2ex}
\section{Proof of Theorem 1}
\label{proof theorem 1}
In order to prove that the problem $\mathcal{P}3$ is a NP-hard problem. We first give a lemma.
\begin{lemma}
    \label{lemma-2}
    The optimal solution to the problem $\mathcal{P}3$ always reaches the equality of the constraints \eqref{p3b}.
\end{lemma}

\begin{IEEEproof}
    We can prove the problem by contradiction. We assume that $\{T^*,\tau_k^*,g_k^*,o_k^*,f_k^{c^*}\}$ is the optimal solution to $\mathcal{P}3$, and we assume that the delay of device $k_{0}$ is smaller than the threshold, i.e.,
    \begin{equation}
        \frac{L_{k_{0}}C_{k_{0}}^{l}}{f_{k_{0}}^{l}}+\frac{L_{k_{0}}D_{k_{0}}o_{k_{0}}^*}{Me^{-g_{k_{0}}^*}}\frac{T_{s}}{\tau_{k_{0}}^*}+\frac{L_{k_{0}}C_{k_{0}}^{e}}{f_{k_{0}}^{c^*}}<T^*. \label{not equal}
    \end{equation}
    We can decrease the $f_{k_{0}}^{c^*}$ to 
    $f_{k_{0}}^{c^{\prime}}$ to make the constraint \eqref{not equal} reach the equality, where $f_{k_{0}}^{c^*}=f_{k_{0}}^{c^{\prime}}+\Delta f$, and the $\Delta f$ can be divided into $K$ parts $\Delta f=\Delta f_{1}+\cdots+\Delta f_{K}$ to allocate all devices, thus the delay of the devices can be given as
    \begin{equation}
        \frac{L_{k}C_{k}^{l}}{f_{k}^{l}}+\frac{L_{k}D_{k}o_{k}^*}{Me^{-g_{k}^*}}\frac{T_{s}}{\tau_{k}^*}+\frac{L_{k}C_{k}^{e}}{f_{k}^{c^*}+\Delta f_{k}}<T^*,\forall k \in \mathcal{K}, k \neq k_{0},
    \end{equation}
    and
    \begin{equation}
        \frac{L_{k}C_{k_{0}}^{l}}{f_{k_{0}}^{l}}+\frac{L_{k_{0}}D_{k_{0}}o_{k_{0}}^*}{Me^{-g_{k_{0}}^*}}\frac{T_{s}}{\tau_{k_{0}}^*}+\frac{L_{k_{0}}C_{k_{0}}^{e}}{f_{k_{0}}^{c^{\prime}}+\Delta f_{k_{0}}}<T^*.
    \end{equation}
    In this case, we can reduce $\tau_{k}^*$ to $\tau_{k}{}^{\prime}$, where $\tau_{k}^*=\Delta\tau_{k}+\tau_{k}{}^{\prime}$ to reach the equality for all devices. Thus, we can further reduce the objective function without violating the constraint, which contradicts the fact that that $T^*$ is the optimal solution. 
\end{IEEEproof}

According to Lemma \ref{lemma-2}, when the solution to $\mathcal{P}3$ is optimal, we have
\begin{equation}
    \frac{L_{k}C_{k}^{l}}{f_{k}^{l}}+\frac{L_{k}D_{k}o_{k}}{Me^{-g_{k}}}\frac{T_{s}}{\tau_{k}}+\frac{L_{k}C_{k}^{e}}{f_{k}^{c}}=T, \forall k \in \mathcal{K}.
\end{equation}
Thus, we obtain $\tau_{k}=a_{k}o_{k}, \forall k \in \mathcal{K},$
where $a_k=\frac{L_kD_ko_kT_s}{Me^{-g_k}}/(T-\frac{L_kC_k^l}{f_k^l}-\frac{L_kC_k^e}{f_k^c}),\forall k \in \mathcal{K}$.

Thus, the problem $\mathcal{P}3$ can be reformulated as
\vspace{-2ex}
\begin{subequations}
    \begin{align}
        \mathcal{P}5: \operatorname*{min}_{\{\tau_{k},g_{k},o_{k},f_{k}^{c}\}}
        & \sum_{k=1}^{K}a_{k}o_{k} \\
        \text{s.t}. 
        & \mathrm{\eqref{p3c}} - \mathrm{\eqref{p3g}}.
    \end{align}
\end{subequations}

We introduce binary variables $x_{k,n} \in \{0, 1\}$, where $x_{k,n}=1$ denotes that the device $k$ selects compression ratio $c_{n}$, and $x_{k,n}=1$ otherwise. Thus, the variable $o_k$ can be denoted as $o_k=\sum_{n=1}^Nc_nx_{k,n}$.
Then, the $\mathcal{P}5$ can be reformulated as
\vspace{-1ex}
\begin{subequations}
    \begin{align}
        \mathcal{P}6: \operatorname*{min}_{\{\tau_{k},g_{k},o_{k},f_{k}^{c}\}}
        & \sum_{k=1}^{K}\sum_{n=1}^N{a_{k}}{c_n}{x_{k,n}} \\
        \text{s.t}. 
        & x_{k,n} \in \{0, 1\}, \forall k \in \mathcal{K}, \forall n \in \mathcal{N}, \\
        & \mathrm{\eqref{p3c}} - \mathrm{\eqref{p3g}}.
    \end{align}
\end{subequations}
It can be observed that $\mathcal{P}6$ can be reduced into a knapsack problem with variables $\{\tau_{k},g_{k},f_{k}^{c}\}$ fixed, which is a NP-hard problem \cite{np-hard}. If $\mathcal{P}6$ can be solved, the knapsack problem can be solved respectively. However, since knapsack problem is NP-hard, and we cannot solve it with polynomial complexity. Therefore, $\mathcal{P}6$ is also an NP-hard problem.
This ends the proof.

\vspace{-2ex}
\section{Proof of Theorem 2}
\label{proof theorem 2}
The objective function and constraints \eqref{p3c} and \eqref{p3d} are linear when $o_k$ is fixed, so we only need to check the convexity of the \eqref{p3b} by computing its Hessian matrix and proving that the Hessian matrix is positive-definite.

We rewrite the $k$-th constraint of \eqref{p3b} as 
\begin{equation}
    f=a+\frac{b}{\tau_{k}}e^{g_{k}}+\frac{c}{f_{k}^{c}}-T,
\end{equation}
where $a=\frac{L_{k}C_{k}^{l}}{f_{k}^{l}}$, $b=\frac{L_{k}D_{k}o_{k}}{Me^{-g_{k}}}\frac{T_{s}}{\tau_{k}}$, $c=\frac{L_{k}C_{k}^{e}}{f_{k}^{c}}$. Thus, the Hessian matrix of $f$ is given as
\begin{equation}
    \begin{aligned}
    \mathbf{H}
    =\begin{bmatrix}\frac b{\tau_k}e^{g_k}&-\frac b{\tau_k^2}e^{g_k}&0\\-\frac b{\tau_k^2}e^{g_k}&2\frac b{\tau_k^3}e^{g_k}&0\\0&0&2\frac c{f_k^{c3}}\end{bmatrix}. \label{Hessian}
    \end{aligned}
\end{equation}


In order to prove that \eqref{p3b} is convex, we can prove that the Hessian matrix in \eqref{Hessian} is positive-definite by proving that all the leading principal minors of $\mathbf{H}$ are positive. We denote the $k$-th order of the leading principal minors of $\mathbf{H}$ by $\left|\mathbf{H}_k\right|$. Then, the first, second, and third order of the leading principal minors of $\mathbf{H}$ are given as $|\mathbf{H}_{1}|=\frac{b}{\tau_{k}}e^{g_{k}}>0$, $|\mathbf{H}_{2}|=\frac{b^{2}}{\tau_{k}{}^{4}}e^{2g_{k}}>0$, $|\mathbf{H}_{3}|=2\frac{c}{{f_{k}^{c}}^{3}}\frac{b^{2}}{\tau_{k}{}^{4}}e^{2g_{k}}>0$, respectively.
It can be seen that all the leading principal minors of $\mathbf{H}$ are positive.
This ends the proof.

\vspace{-2ex}
\section{Proof of Theorem 3}
\label{proof theorem 3}
    According to the KKT conditions, the necessary and sufficient conditions are given as
    \begin{equation}
        \frac{\partial\mathcal{L}}{\partial\tau_k^*}=1-\lambda_k^*\frac{L_kD_ko_k}M\frac{T_s}{\tau_k^*}e^{g_k^*}=
        \begin{cases}=0,\tau_k^*\geq0,\\
        \geq0,\tau_k^*=0,\end{cases} \forall k, \label{grad_tauk}
    \end{equation}
    \begin{equation}
        \frac{\partial\mathcal{L}}{\partial f_k^{c^*}}=-\lambda_k^*\frac{L_kC_k^e}{f_k^{c^*2}}+\mu^*=
        \begin{cases}=0,f_k^{c^*}\geq0,\\
        \geq0,f_k^{c^*}=0,\end{cases} \forall k, \label{grad_fk}
    \end{equation}
    \begin{equation}
        \frac{\partial\mathcal{L}}{\partial g_k^*}=\lambda_k^*\frac{L_kD_ko_k}M\frac{T_s}{\tau_k^*}e^{g_k^*}=
        \begin{cases}=0,g_k^*\geq d_k,\\
        \geq0,g_k^*=d_k,\end{cases} \forall k, \label{grad_gk}
    \end{equation}
    \begin{equation}
        \lambda_k^*\left(\frac{L_kC_k^l}{f_k^l}+\frac{L_kD_ko_k}{M}\frac{T_s}{\tau_k^*}e^{g_k^*}+\frac{L_kC_k^e}{f_k^{c^*}}-T\right)=0,\forall k, \label{kkt com1}
    \end{equation}
    \begin{equation}
        \mu^*(\sum_{k=1}^Kf_k^{c^*}-F^c)=0, \mu^{*}\geq0 \label{kkt com2}
    \end{equation}
    \begin{equation}
        \lambda_{k}^{*}\geq0,\forall k.
    \end{equation}

    For constraint \eqref{p3b}, we can find that $\tau_k^*>0$ and $f_k^{c^*}>0$. Thus, we can derive that $\lambda_k^*>0$ and $\mu^*>0$ according to \eqref{grad_tauk} and \eqref{grad_fk}, respectively. Since $\lambda_k^*>0$, it is obvious that $g_k^*=d_k$ according to \eqref{grad_gk}. Moreover, combining \eqref{grad_tauk} and \eqref{grad_fk}, we can derive the relationship between $\tau_k^*$ and $f_k^{c^*}$, as
    \begin{equation}
        f_k^{c^*}=\tau_k^*\sqrt{\frac{MC_k^e}{\mu^*D_ko_ke^{g_k^*}T_s}}, \forall k.
    \end{equation}
    Next, according to \eqref{kkt com1}, we have
    \begin{equation}
        f_{k}^{c}{}^{*}=\frac{L_{k}}{T-\frac{L_{k}C_{k}^{l}}{f_{k}^{l}}}(\sqrt{\frac{D_{k}o_{k}e^{g_{k}^{*}}T_{s}C_{k}^{e}}{M\mu^{*}}}+C_{k}^{e}),
    \end{equation}
    where $\mu^*$ satisfy $\sum_{k=1}^Kf_k^{c^*}-F^c=0$. Thus, we can obtain
    \begin{equation}
    \mu^*=\left(\frac{\sum_k\frac{L_k}{T-{L_kC_k^l}/{f_k^l}}\sqrt{\frac{D_{0}o_ke^{g_k^*}T_sC_k^d}{M}}}{F^c-\sum_k\frac{L_kC_k^d}{T-{L_kC_k^l}/{f_k^l}}}\right)^2,
    \end{equation}
    which ends the proof.
\end{appendices}


\vspace{10pt}


\begin{thebibliography}{99}
    \scriptsize
    \bibitem{shannon} T. M. Cover and J. A. Thomas, ``Information theory and statistics,'' \textit{Elements Inf. Theory}, vol. 1, no. 1, pp. 279--335, 1991.
    
    \bibitem{deepsc} H. Xie, Z. Qin, G. Y. Li, and B.-H. Juang, ``Deep learning enabled semantic communication systems,'' \textit{IEEE Trans. Signal Process.}, vol. 69, pp. 2663--2675, Apr. 2021.
    
    \bibitem{jscc} E. Bourtsoulatze, D. B. Kurka, and D. G\"{u}nd\"{u}z, ``Deep joint source-channel coding for wireless image transmission,'' \textit{IEEE Trans. Cogn. Commun. Netw.}, vol. 5, no. 3, pp. 567--579, May 2019.
    
    \bibitem{speech} T. Han, Q. Yang, Z. Shi, S. He, and Z. Zhang, ``Semantic-preserved communication system for highly efficient speech transmission,'' \textit{IEEE J. Sel. Areas Commun.}, vol. 41, no. 1, pp. 245--259, Jan. 2023.
    
    \bibitem{performance} Y. Wang, M. Chen, W. Saad, T. Luo, S. Cui, and H. V. Poor, ``Performance optimization for semantic communications: An attention-based reinforcement learning approach,'' \textit{IEEE J. Sel. Areas Commun.}, vol. 40, no. 9, pp. 2598--2613, Sep. 2022.

    \bibitem{resource} L. Yan, Z. Qin, R. Zhang, Y. Li, and G. Y. Li, ``Resource allocation for text semantic communications,'' \textit{IEEE Wireless Commun. Lett.}, pp. 1394--1398, Apr. 2022.

    \bibitem{NOMA} X. Mu, Y. Liu, L. Guo, and N. Al-Dhahir, ``Heterogeneous semantic and bit communications: A semi-NOMA scheme,'' \textit{IEEE J. Sel. Areas Commun.}, vol. 41, no. 1, pp. 155--169, Jan. 2023.
    
    
    

    \bibitem{cnn} Y. He, J. Ren, G. Yu, and Y. Cai, ``Optimizing the learning performance in mobile augmented reality systems with CNN,'' \textit{IEEE Trans. Wireless Commun.}, vol. 19, no. 8, pp. 5333--5344, Aug. 2020.
    
    \bibitem{over-the-air} G. Zhu, Y. Wang, and K. Huang, ``Broadband analog aggregation for low-latency federated edge learning,'' \textit{IEEE Trans. Wireless Commun.}, vol. 19, no. 1, pp. 491--506, Jan. 2020.

    \bibitem{capacity optimization} K. Chi, Q. Yang, Z. Yang, Y. Duan, and Z. Zhang, ``Resource allocation for capacity optimization in joint source-channel coding systems,'' in \textit{Proc. IEEE Int. Conf. Commun. (ICC)}, 2023, pp.2099--2104.

    \bibitem{delay} Z. Yang, M. Chen, W. Saad, C. S. Hong, and M. Shikh-Bahaei, ``Delay minimization for federated learning over wireless communication networks,'' in \textit{Proc. Int. Conf. Mach. Learn. Workshop}, Jul. 2020, pp. 1--7.

    \bibitem{imagenet} O. Russakovsky et al., ``ImageNet large scale visual recognition challenge,'' \textit{Int. J. Comput. Vis.}, vol. 115, no. 3, pp. 211--252, Dec. 2015.

    \bibitem{cpulimit} A. Marletta. (2012). \textit{Cpulimit}. [Online]. Available: https://github.com/opsengine/cpulimit.

    \bibitem{np-hard} J. Kleinberg and E. Tardos, \textit{Algorithm Design}. Pearson Education India, 2006.
\end{thebibliography}
	
\end{document}